\documentclass[a4paper,12pt]{article}
\usepackage{graphics}
\usepackage{amsfonts}
\usepackage{epsfig,amssymb,amsmath,graphicx,caption,subcaption,
verbatim,hyperref,xcolor,ulem,epstopdf,psfrag,pstool,braket,array,
enumerate,wrapfig,tabularx}

\begin{document}
%

\newcommand{\be}{\begin{equation}}
\newcommand{\ee}{\end{equation}}
\newcommand{\bea}{\begin{eqnarray}}
\newcommand{\eea}{\end{eqnarray}}
\newcommand{\nn}{\nonumber}                  
\newcommand{\bean}{\begin{eqnarray*}}
\newcommand{\eean}{\end{eqnarray*}}
\font\upright=cmu10 scaled\magstep1
\font\sans=cmss12
\newcommand{\ssf}{\sans}
\newcommand{\stroke}{\vrule height8pt width0.4pt depth-0.1pt}
\newcommand{\Z}{\mathbb{Z}}
\newcommand{\ZZ}{\Z\hskip -10pt \Z_2}
\newcommand{\C}{{\rlap{\upright\rlap{C}\kern 3.8pt\stroke}\phantom{C}}}
\newcommand{\R}{\hbox{\upright\rlap{I}\kern 1.7pt R}}
\newcommand{\HH}{\hbox{\upright\rlap{I}\kern 1.7pt H}}
\newcommand{\CP}{\hbox{\C{\upright\rlap{I}\kern 1.5pt P}}}
\newcommand{\identity}{{\upright\rlap{1}\kern 2.0pt 1}}
\newcommand{\half}{\frac{1}{2}}
\newcommand{\quart}{\frac{1}{4}}
\newcommand{\pr}{\partial}
\newcommand{\bm}{\boldmath}
\newcommand{\I}{{\cal I}} 
\newcommand{\M}{{\cal M}}
\newcommand{\N}{{\cal N}}
\newcommand{\e}{\varepsilon}

\thispagestyle{empty}
\vskip 3em
\begin{center}
{{\bf \Large Wormhole Model for Neon-20 
}} 
\\[15mm]

{\bf \large Nicholas~S. Manton\footnote{email: N.S.Manton@damtp.cam.ac.uk}
and Maciej Dunajski\footnote{email: M.Dunajski@damtp.cam.ac.uk}}\\[20pt]

\vskip 1em
{\it 
Department of Applied Mathematics and Theoretical Physics,\\
University of Cambridge, \\
Wilberforce Road, Cambridge CB3 0WA, U.K.}
\vspace{12mm}

\abstract
{A quantum mechanical model for the Neon-20 nucleus is developed that
allows for the splitting of a bipyramidal structure of five
alpha-partices into an alpha-particle and an Oxygen-16 nucleus. The
geometry of the configuration space is assumed to be a 3-dimensional
spatial wormhole, and on the wormhole background there is an attractive
short-range potential. This leads to a radial Schr\"odinger equation
of the Heun form, which simplifies for threshold bound states to an
associated Legendre equation that has explicit solutions. The energies
of the true bound states for all spin/parities are numerically calculated,
and match those of the well-established $K^\pi=0^+$ and $K^\pi=0^-$
rotational bands of Neon-20, and certain higher bands.}
\end{center}

\vskip 150pt
\leftline{Keywords: Spatial Wormhole, Neon-20 Nucleus, Rotational Bands}
\vskip 1em

\vfill
\newpage
\setcounter{page}{1}
\renewcommand{\thefootnote}{\arabic{footnote}}


\section{Motivation} 
\vspace{4mm}

We develop a model for quantum states of the Neon-20 nucleus, where the
underlying configuration space is a spatial, spherically-symmetric wormhole.
Spacetime wormholes have generated much interest, but are usually regarded
as a science-fiction fantasy. Here we find a wormhole that is physically
realised -- not as a spacetime, but as a model configuration space for
a spatially-extended, deformable nucleus. The characteristic
length scale of the wormhole is about 5 fm, comparable to the
size of the Neon-20 nucleus. This wormhole model was first outlined
in \cite{Emerging}, but not analysed there in any detail.

We start from the old idea that the ground-state configuration of
Neon-20 is a bipyramid of five alpha-particles \cite{Wheel,Wef,HT}, which
splits relatively easily into a single alpha-particle and a bound
cluster of four alpha-particles representing Oxygen-16. The
bipyramid has an equilateral triangle of alphas in the middle, but we
ignore rotations in the plane of this triangle and treat the bipyramid
as axially symmetric. The angular momentum projection about
the symmetry axis is therefore $K=0$. We also neglect centre of mass motion.

Conventionally, one would have a Euclidean configuration space for the
separation vector of the two clusters. A central ball can then be cut out to
enforce a non-zero minimal separation, or a potential can be introduced
that strongly disfavours the clusters from overlapping. Such models
have rather singular behaviour near the bipyramid.

In contrast, our model has a smooth, curved configuration space near
the bipyramid, avoiding any singularity. It has
a single radial coordinate $r$, which accurately corresponds to the
separation of the clusters asymptotically, when they are
well-separated, and additionally it has the usual spherical polar
coordinates $(\theta,\phi)$ representing the direction of the
separation axis. Unusually, however, $r$ runs from $-\infty$
to $\infty$. The metric is that of an Ellis--Bronnikov wormhole \cite{Ell,Bro},
\be
ds^2 = dr^2 + (r^2 + a^2)(d\theta^2 + \sin^2 \theta \, d\phi^2),
\label{metric}
\ee
where the radius of the wormhole's throat, $a$, is related to the
linear size of the bipyramid.

On the wormhole background, there is a smooth potential energy
\[
-\frac{V_0}{(r^2+a^2)^2} \,,
\]
with $V_0$ a positive constant. This potential depends symmetrically
on $r$, and its minimum at $r=0$ is where the clusters merge into the
bipyramid. There is more than one motivation for this choice of potential.
It is attractive, and short-ranged compared to the centrifugal
repulsion that occurs in the presence of non--zero angular momentum.
It is also a multiple of the Ricci scalar curvature of the wormhole,
something that could arise from an alternative treatment of canonical
quantization. Finally, the potential's algebraic form means that at
the threshold energy for bound states, the radial Schr\"odinger equation
usefully simplifies to an associated Legendre equation, which can be
explicitly solved. 

\begin{center}
\includegraphics[scale=0.2,angle=0]{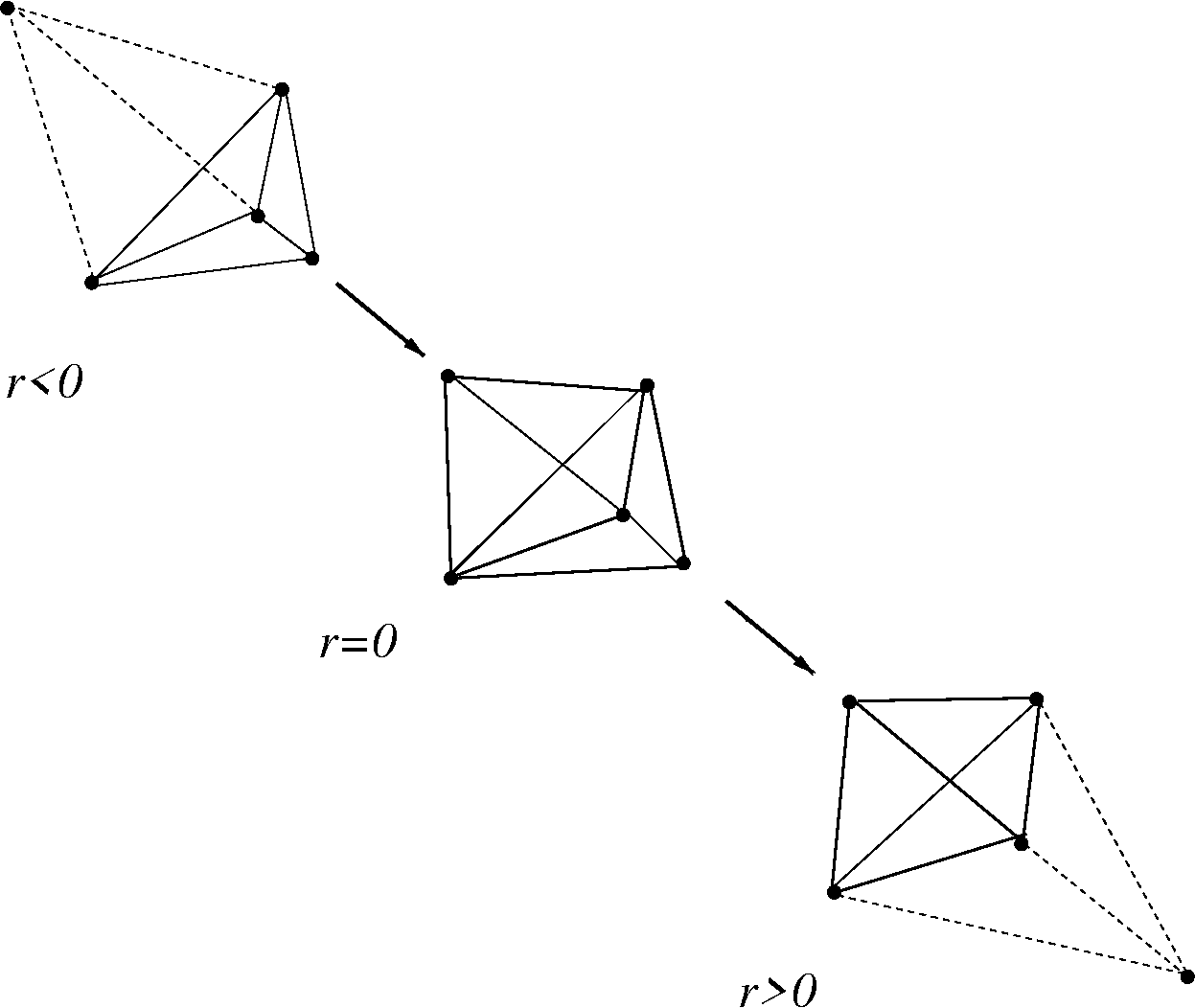}
\begin{center}
{{\bf Figure 1.} {An incoming alpha-particle approaches a
tetrahedron of alpha-particles, forms a bipyramid instantaneously,
and then the opposite alpha-particle is ejected.}}
\end{center}
\end{center}

To justify the wormhole geometry, consider a simple classical motion with
$r$ monotonically decreasing through zero, and $(\theta,\phi)$ fixed.
This models an incoming alpha-particle coalescing instantaneously with
the four-alpha cluster to form the bipyramid, and a different
alpha-particle from the opposite side of the bipyramid getting
ejected (see Fig.1). The motion can be thought of as
analogous to that of a Newton cradle, with one ball coming in and a
different ball going out. If the incoming and outgoing
alpha-particles are distinguishable, it makes sense for $r$ to have
the range $(-\infty,\infty)$. However, note that the outgoing
particle's angular coordinates are antipodal to those of the
incoming particle, and because alpha-particles are indistinguishable
bosons, one should identify points in configuration space with
coordinates $(r,\theta,\phi)$ and $(-r, \pi-\theta, \phi + \pi)$. The
true configuration space is therefore not the complete wormhole, but
its quotient by $\Z_2$ under this identification.

The $\Z_2$ action has no fixed points, so the quotient space is
smooth. The quotient is the half-wormhole $r \ge 0$, where
antipodal points $(\theta,\phi)$ and $(\pi-\theta, \phi + \pi)$ on the
2-sphere at $r=0$ are identified. This identification is
required, because the bipyramid is symmetric under the antipodal map.
We will quantize the dynamics on the half-wormhole, but to do this
it is convenient to perform the quantization on the complete wormhole,
separating radial and angular variables as usual; the $\Z_2$
symmetry then constrains physical states to be either symmetric in $r$
with even angular momentum, or antisymmetric in $r$ with odd angular
momentum. A state's parity is even/odd if the angular momentum is even/odd.

This wormhole model is an anharmonic extension of a rovibrational
model for Neon-20, incorporating only the lowest-frequency vibrational
mode of the bipyramid, the singly-degenerate mode transforming
under the $A_2''$ representation of ${\cal D}_{3h}$ that tends
to produce a $1+4$ cluster split. The mode's oscillating amplitude
corresponds to an oscillation of $r$ around zero. The potential on the
wormhole is minimal at the bipyramid, just like the harmonic oscillator
potential. Another similarity is that the quantized rovibrational model has
a bosonic constraint. Physical states are those with either an even
number of vibrational phonons and even angular momenta, or an odd number of
phonons and odd angular momenta.

Apart from the wormhole radius $a$, our model has just one adjustable,
dimensionless parameter $m$. It can take any positive value, and we
found initially that a good fit to experimental data was with
$m=9.01$. Mathematically, it is very convenient for $m$ to be a
positive integer, so we have fixed $m=9$. A consequence is that all
the threshold bound states, just below the continuum scattering
states, have radial wavefunctions given by
analytic solutions of an associated Legendre equation.

Having fixed $m=9$, together with an energy scale, we obtain a good match to
the lowest observed rotational bands of Neon-20 with quantum numbers
$K^\pi=0^+$ and $K^\pi=0^-$, interpreted here as zero-phonon and one-phonon
bands. Bijker and Iachello have presented a rovibrational analysis of
Neon-20 states, incorporating all modes of small oscillation of the bipyramid
\cite{BI}. Our interpretation of some of the higher $K=0$ bands differs
somewhat from theirs. We agree on the two-phonon $0^+$ band, but
interpret the second $0^-$ band as a three-phonon band, rather than
a band combining two distinct phonons with opposite $\Z_2$ symmetries.
Quantitatively, our results are comparable, but as we have fewer
adjustable parameters, we can predict some of the parameters that
they have fitted -- in particular, the ratios of the rotational
constants $B$ associated to the various $K=0$ bands.

\vspace{4mm}

\section{Quantum States on the Wormhole}
\vspace{3mm}

We start with the classical Lagrangian dynamics for the relative
motion of an alpha-particle and a four-alpha cluster, and then
quantize. The Lagrangian is
\be
L = \half \mu \left({\dot r}^2 + (r^2 + a^2)({\dot\theta}^2
+ \sin^2\theta \, {\dot\phi}^2)\right) + \frac{V_0}{(r^2+a^2)^2} \,,
\label{Lag}
\ee
where the kinetic term is based on the wormhole metric, $\mu$ is the
reduced mass of the two clusters at large separation, and
the potential is negative and attractive. The physical interpretation
of the wormhole geometry is that the inertial mass tensor is
separation-dependent, and different in the radial and angular
directions. A separation-dependent inertial mass in the radial
direction has been previously considered in the context of Neon-20 by
Wen and Nakatsukasa \cite{WN}.

The quantum Hamiltonian derived from $L$ is
\bea
H &=& -\frac{\hbar^2}{2\mu} \nabla^2 - \frac{V_0}{(r^2+a^2)^2} \nn \\
&=& -\frac{\hbar^2}{2\mu} \left( \pr_{rr} + \frac{2r}{r^2+a^2} \, \pr_r
  + \frac{1}{r^2+a^2} \nabla_{\rm ang}^2 \right) - \frac{V_0}{(r^2+a^2)^2} \,,
\eea
where $\nabla^2$ is the Laplace--Beltrami operator for the wormhole
metric (\ref{metric}), and $\nabla_{\rm ang}^2$ is the usual
angular part of the Laplacian in spherical polars. $\nabla^2$ reduces
to the standard Laplacian in spherical polars when $a=0$.
We separate variables, and write stationary state wavefunctions as
\be
\Phi(r,\theta,\phi) = \chi(r)Y_{l,\widetilde{m}}(\theta,\phi) \,.
\ee
$l$ is the usual angular momentum quantum number of the spherical
harmonic $Y_{l,\widetilde{m}}$, and the
index $\widetilde{m}$ takes its usual $2l+1$ values. We are reserving $m$
to be a parameter in the potential. The reduced, radial equation for
stationary states of energy $E$ is
\be
-\frac{\hbar^2}{2\mu} \left( \frac{d^2 \chi}{dr^2}
+ \frac{2r}{r^2+a^2} \frac{d\chi}{dr}
- \frac{l(l+1)}{r^2+a^2} \, \chi \right) - \frac{V_0}{(r^2+a^2)^2} \,
\chi = E \chi \,.
\ee
We recall that the bound states of this potential with zero angular
momentum have, in the context of kinks on the wormhole background,
been considered in \cite{BDKK} and \cite{Alice}.

We now change variable by setting $r = ax$, and write $V_0 =
\frac{\hbar^2 a^2}{2\mu} m^2$ and $E = \frac{\hbar^2}{2\mu a^2} \e$.
The radial equation then has the dimensionless form
\be
-\frac{d^2 \chi}{dx^2} - \frac{2x}{x^2+1} \frac{d\chi}{dx}
+ \frac{l(l+1)}{x^2+1} \, \chi - \frac{m^2}{(x^2+1)^2} \, \chi
= \e \chi \,.
\label{radial}
\ee
Equation (\ref{radial}) for general $\e$ is a confluent Heun equation,
with two regular singular points at $x = \pm i$, and a confluent
singularity at infinity. Its general solution is given by
\begin{eqnarray}
& \chi(x) = c_1\left( {x}^{2}+1 \right)^{{\frac {m}{2}}}
{\mbox{HeunC}} \left(0,-{\frac{1}{2}},m,-{\frac {\e}{4}},
{\frac{m^2 + 1 - l(l+1) + \e}{4}},-{x}^{2} \right) \nonumber \\
& + \, c_2 \, x \left( {x}^{2}+1 \right) ^{{\frac{m}{2}}}
{\mbox{HeunC}} \left( 0,{\frac{1}{2}},m,-{\frac {\e}{4}},
{\frac{m^2 + 1 - l(l+1) + \e}{4}},-{x}^{2} \right). 
\end{eqnarray}

\vspace{4mm}

\section{The Energy Spectrum}
\vspace{3mm}

Here we derive some general properties of the bound state spectrum of
the radial equation (\ref{radial}). True bound states are solutions
with negative $\e$, but there are also normalisable, threshold bound
states with $\e = 0$.

If we write $\chi(x) = (x^2 + 1)^{-\half} \, \eta(x)$, eq.(\ref{radial}) becomes
\be
-\frac{d^2 \eta}{dx^2} + \frac{l(l+1)}{x^2+1} \, \eta
- \frac{m^2 - 1}{(x^2+1)^2} \, \eta = \e \eta \,,
\label{1deq}
\ee
which is easier to treat analytically. The transformed equation
(\ref{1deq}) is a 1-d Schr\"odinger equation with effective potential (Fig.2)
\be
v_{\rm eff}(x) = \frac{l(l+1)}{x^2+1} - \frac{m^2 - 1}{(x^2+1)^2} \,.
\label{veff}
\ee
\begin{center}
\includegraphics[scale=0.6,angle=0]{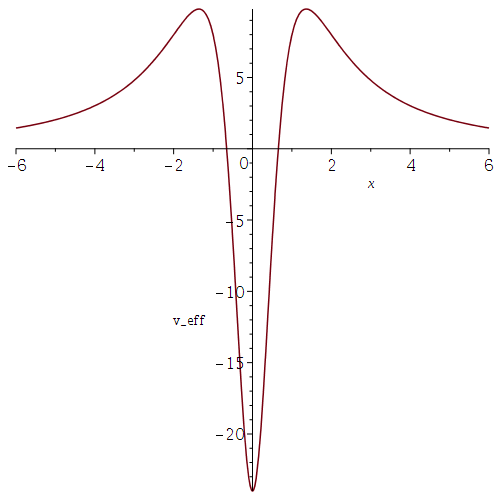}
\begin{center}
{{\bf Figure 2.} {Effective potential with $m=9$ and $l=7$. The
centrifugal repulsion dominates the attractive potential at large
cluster separation.}}
\end{center}
\end{center}
For bound states to exist, $v_{\rm eff}$ has to be negative
somewhere, which requires $l(l+1) < m^2 - 1$. On the other hand, if
$2l(l+1) < m^2 - 1$ then the integral of $v_{\rm eff}$ over $\R$ is negative,
and a bound state definitely exists (by a variational argument). For
general positive $m$ we therefore expect bound states, but only for $l < m$.

Rather remarkably, there are threshold bound states for all $0 \le l <
m$ when $m$ is a positive integer. To see this, we rewrite eq.(\ref{radial})
for $\e =0$ as
\be
(x^2+1)\frac{d^2 \chi}{dx^2} + 2x \frac{d\chi}{dx}
- l(l+1) \chi + \frac{m^2}{x^2+1} \, \chi = 0 \,.
\label{zeroenergy}
\ee
Setting $z=ix$, this becomes the standard associated Legendre
equation. We are interested in solutions for real $x$ that are
normalisable, and these are associated Legendre functions evaluated on the
imaginary $z$-axis. As this axis does not pass through the
regular singular points at $z = \pm 1$, the solutions can be
singular at these points, and we are not constrained to impose $|m| \le l$
as in the construction of spherical harmonics. 

We can therefore assume that $\chi$ has the form
\be
\chi(x) = \frac{S(x)}{(x^2 + 1)^{\frac{m}{2}}}
\ee
where
\be
S(x) = \sum_{k=0} a_k \, x^k \,.
\ee
From (\ref{zeroenergy}) we then derive the recurrence relation
\be
a_{k+2} = - \frac{m(m-1) - l(l+1) - (2m-1)k + k^2}{(k+2)(k+1)} \,
a_k \,. 
\ee
$\chi(x)$ is the radial wavefunction of a normalisable, threshold bound
state provided the series for $S$ truncates to a finite polynomial
of degree less than $m-1$. This occurs for all angular momenta in the
range $0 < l < m$, and we denote the resulting polynomial
$S_l^m(x)$. $S_l^m$ has degree $m-l-1$; it is even if $m-l-1$ is
even and odd if $m-l-1$ is odd, and has $m-l-1$
nodes. A further threshold state exists for $l=0$, but this is not
normalisable because $S_0^m$ has degree $m-1$. Note that all the
normalisable threshold states are physically allowed provided $m$ is
odd, because $m-l-1$ has the same parity as $l$ in this case.

As explained in Sect.1, we have made the choice $m=9$. Usefully,
for fitting the Neon-20 spectrum, there are then threshold states for all
angular momenta up to $l=8$. For $m=9$, and $l$ decreasing from $8$ to $0$
the polynomials $S_l^9(x)$ are respectively
\begin{eqnarray}
&1, \;x,\; 1-15x^2,\; 3x-13x^3,\; 3-78x^2+143x^4,\; 15x-110x^3+99x^5,
\nonumber \\
&5-165x^2+495x^4-231x^6,\; 5x-45x^3+63x^5-15x^7, \nonumber \\
&1-36x^2+126x^4-84x^6+9x^8,
\end{eqnarray}
and all their roots are real.  

$\chi(x)$ is in fact, up to a normalisation constant, the associated
Legendre function of the second kind $Q_l^m(z)$, evaluated on the
imaginary axis $z=ix$. It has the explicit form, for $m$ an integer
greater than $l$,
\be
Q_l^m(z) = (1 - z^2)^{\frac{m}{2}} \, \frac{d^m}{dz^m}
\left[ \ln \left( \frac{1+z}{1-z} \right)
\frac{d^l}{dz^l} (1 - z^2)^l  \right] \,.
\ee
One might anticipate, from the expression for $Q_l(z)$ in ref.\cite{DigL},
an additional polynomial inside
the square brackets, but its $m$th derivative vanishes. Also, because
$m > l$, at least one derivative acts on the logarithmic term, so
there is no such term in the result. 

Let us now consider the negative energy bound states. Because the
threshold bound state with angular momentum $l$ has
$m-l-1$ nodes, it follows from the Sturm oscillation theorem that there are
$m-l-1$ negative energy states with angular momentum $l$. These
are confluent Heun functions, but as there are no simple formulae for
their energies, we have solved eq.(\ref{1deq}) numerically
to find them. We use a shooting and bisection method on the interval
$[-20, 20]$, using as initial condition at $x=0$ either that $\eta$
has zero slope (for the even solutions) or zero value and a finite
slope (for the odd solutions).

For each $l$, the lowest-energy wavefunction $\eta(x)$ is symmetric
in $x$ and has no nodes, but as the energy increases, the parity of
$\eta$ alternates and the number of nodes increases by 1. The physically
allowed states are those where the parity of $\eta$ matches the parity
of $l$. The spectrum of physically allowed states for $m=9$, including
threshold states, is shown in Fig.3. 
\begin{center}
\includegraphics[scale=0.6,angle=0]{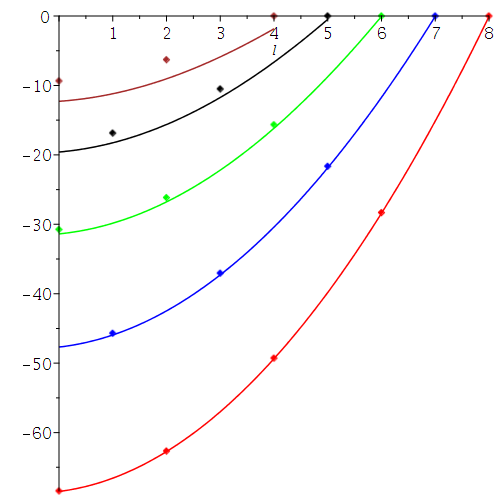}
\begin{center}
{{\bf Figure 3.} {Numerical bound state spectrum for $m=9$ and integer
$l < 9$ (dots), compared with the anharmonic oscillator
approximation (curves).}}
\end{center}
\end{center}

Although the true energies of the bound states are not known
analytically, they can be evaluated using an anharmonic oscillator
approximation. The expansion of $v_{\rm eff}$ up to $O(x^4)$ is
\be
v_{\rm eff} \simeq -(m^2 - 1) + l(l+1) + [2(m^2-1) - l(l+1)]x^2
- [3(m^2-1) - l(l+1)]x^4 \,.
\label{veffaho}
\ee
Using the harmonic oscillator wavefunctions for $v_{\rm eff}$
truncated at order $x^2$, together with first-order perturbation
theory to take account of the $x^4$ term \cite{LL}, we find the
approximate energy levels 
\bea
\e_{n,l} &=& -(m^2 - 1) + l(l+1) + (2n+1)\sqrt{2(m^2-1) - l(l+1)}
\nonumber \\
&& \qquad - \frac{3}{2} \, \frac{3(m^2-1) - l(l+1)}{2(m^2-1) - l(l+1)}
\left(n^2 + n + \half \right) \,,
\eea
where $n \ge 0$ is the harmonic oscillator level. These approximate
energies are also shown in Fig.3 for $m=9$, as functions
of continuous $l$. The lowest-lying states for each even $l$
comprise the ground state rotational band, with $n=0$, and above this
we see the higher rotational bands. The $n$th rotational band can be
interpreted as rotational excitations of an $n$-phonon, purely
vibrational state.

For $m=9$, the anharmonic oscillator energies are accurate only up to
about $n=4$, but the phonon number $n$ is still a good label for all
the bound states. The rotational bands get shorter as $n$ increases,
because the threshold state is reached when $l = 8-n$.

In each rotational band, we can express the energy for small $l$ as
\be
\e_{n,l} \simeq \e_n + b_n \, l(l+1) \,,
\ee
where $b_n$ is the rotational constant (inversely proportional to the
effective moment of inertia of the band). The anharmonic oscillator
approximation gives, for $m=9$,
\be
b_n \simeq 1 - \frac{1}{\sqrt{640}} (2n+1)
- \frac{3}{640} \left(n^2 + n + \half \right) \,.
\label{rotconst}
\ee
The values for the lowest few bands are $b_0 = 0.96 \,, b_1 = 0.86 \,, b_2 =
0.77 \,, b_3 = 0.66$ and $b_4 = 0.54$. We see from Fig.3 that these
are quite accurate, even though the actual rotational bands are
significantly shifted up for $n=3$ and $n=4$.

A selection of bound state wavefunctions $\eta(x)$ is shown in
Figs.4a, 4b, 4c.
\begin{center}
\includegraphics[scale=0.5,angle=0]{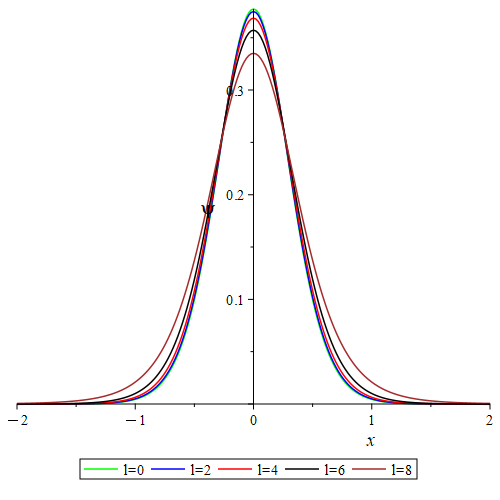}
\begin{center}
{{\bf Figure 4a.} {Bound states with $n=0$.}}
\end{center}
\end{center}
\begin{center}
\includegraphics[scale=0.5,angle=0]{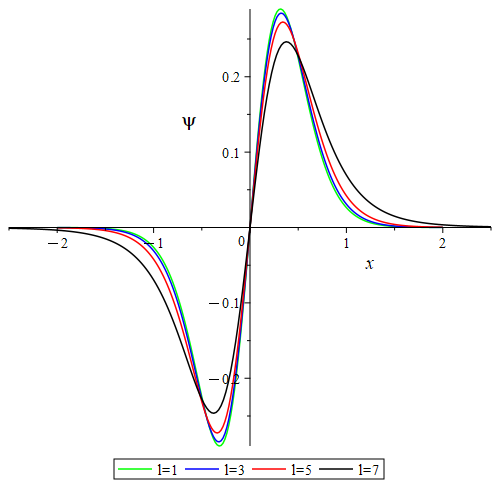}
\begin{center}
{{\bf Figure 4b.} {Bound states with  $n=1$.}}
\end{center}
\end{center}
\begin{center}
\includegraphics[scale=0.5,angle=0]{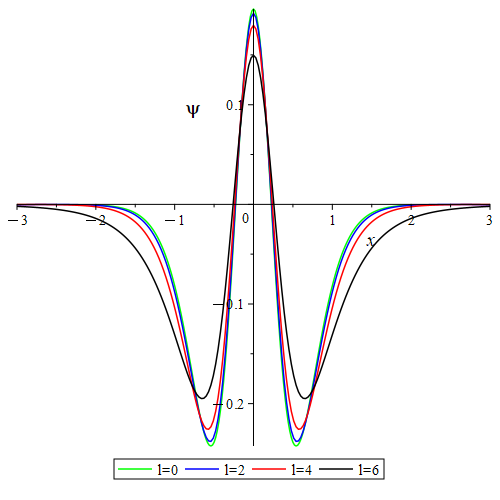}
\begin{center}
{{\bf Figure 4c.} {Bound states with $n=2$.}}
\end{center}
\end{center}
For given $n$, and independently of $l$, they have $n$ nodes
in the full range $-\infty < x < \infty$ and their widths hardly
change with $l$ because they are well approximated by harmonic
oscillator wavefunctions in the simplified, $l$-independent effective
potential derived from (\ref{veff}),
\be
\widetilde{v_{\rm eff}} = -m^2 + 2m^2 x^2 \,.
\ee
However, the threshold states, with maximal $l$, are broader and decay
more slowly for large $|x|$.

These wavefunctions may be compared to those in refs.\cite{BDV} and
\cite{Fuj}, whose node number varies with $l$. However, most of the
latter nodes are in the cluster-overlap region, and the wormhole
wavefunctions are similar in shape only if the comparison is made
outside this region. Our wormhole model has some analogy to
the pseudopotential model for the states of outer valence electrons
in atoms \cite{pseudo}. Wavefunctions in pseudopotentials are suppressed
in the atomic interior and have a reduced number of radial nodes.
However, in our model of Neon 20, there is a change in the nuclear
inter-cluster geometry, and not just in the potential.

\vspace{4mm}

\section{Comparison with Neon-20 States}
\vspace{3mm}

For the experimental spectrum of Neon-20, we use the TUNL
tables \cite{TUNL}; see also the review \cite{Tilley} and the
ENSDF tables \cite{ENSDF}.
The identification of rotational bands in Neon-20 goes
back many decades. See, for example, refs.\cite{Lit, Bou, Fuj, CL, Mar}
among many others, also Table 20.20 in \cite{Tilley}, and the
recent discussion by Bijker and Iachello \cite{BI}. The ground state $0_1^+$
rotational band is well established up to $l=6$. There is also a
well-established, lowest $0^-$ rotational band with states from $l=1$
up to $l=7$. We identify these bands with the 0-phonon
and 1-phonon bands in our model. There are a few more recognised $0^+$
bands. Following Bijker and Iachello, we regard the $0_2^+$ and
$0_3^+$ bands as arising from other, symmetric vibrational excitations of the
bipyramid, and identify the experimental $0^+_4$ band as the
2-phonon band in our model. The states of the $0^+_4$ band have been
recognised as ``higher nodal'' states by Fujiwara et al. \cite{Fuj},
and our model gives them the expected
wavefunction structure, with a single node in the half-wormhole range 
$x > 0$. We have also tentatively identified one higher, observed rotational
band with the 3-phonon band of our model. A few further observed
states can be assigned to the shorter bands with 4-, 5-, 6- and 7-phonons.
$0^+$ bands have positive-parity states of even angular momenta, and
$0^-$ bands have negative-parity states of odd angular momenta.

To calibrate our model, we need to shift the model ground state
upwards to match the experimental energy $E=0$, and then find the
optimal physical energy scale $\hbar^2/2\mu a^2$ by a least squares fit.
The model's dimensionless ground state energy is $\e = -68.5$
(both numerically and in the anharmonic oscillator approximation)
so the shift upwards is by $68.5$.

In a preliminary fit we matched the states that are most confidently
assigned to rotational bands -- those in the ground state $0^+_1$
band up to $6^+$, in the lowest $0^-$ band up to $7^-$,
and in the $0^+_4$ band up to $4^+$. Following refs.\cite{Fuj,Mar}
and others, we identified the $7^-$ state at 15.37 MeV to be in the
$0^-$ band; it is a threshold state in our model. Consistently
with this, the preliminary fit gave a threshold energy just above 15 MeV.
We then noted that the states in the TUNL table between 15.1 MeV and
15.9 MeV are unique candidates for all the threshold states in our
model, with spin/parities between $1^-$ and $8^+$. All the required
spin/parities occur, and there are almost no other observed states in this
energy range, one exception being the $8^-$ state assigned to the $2^-$ band of
Neon-20.

For our final calibration, we therefore extended our ground state
rotational band to include the 15.87 MeV $8^+$ state as the threshold
state. Other models also predict that the $8^+$ state in the ground state
band has a similar energy \cite{BDV}. There is one lower $8^+$ state at 11.95
MeV that is often assigned to the ground state band, but this creates a sharp
kink in the band slope, and the state has other properties that makes
this assignment controversial, for example, its rather low E2 transition
rate to the $6^+$ state in the ground state band \cite{Tilley}. We also
followed Michel et al. \cite{MI} in assigning the 15.16 MeV $6^+$ state to
the observed $0^+_4$ band of broad, higher nodal states. They argued that
the energy of the $6^+$ state in this band is in the range 14-15 MeV, rather
than at 12.58 MeV. It becomes the threshold state in our 2-phonon band.
The broad $5^-$ state at 15.17 MeV becomes the threshold state in the 3-phonon
band. We have also identified various lower-spin states in the TUNL
table that are close in energy to the states required for our model,
both at and below the threshold, but these states are more numerous,
so we have some choice, and they have less significance.

In the best fit of our model to all these 24 identified states, the
conversion factor from the shifted $\e$ to the physical energy $E$
is $\hbar^2/2\mu a^2 = 0.222$ MeV. The threshold energy is then at
$E_{\rm thresh} = 68.5 \times 0.222 \, {\rm MeV} = 15.19$ MeV.
Setting $\hbar = 197.3$ MeV fm, and $\mu = 2982$ MeV (the reduced mass
$\mu$ being almost exactly four-fifths of the alpha-particle mass
$m_\alpha$), we find $a = 5.42$ fm. This is the radius of the 2-sphere
throat of the wormhole, and has order of magnitude the linear size of the
bipyramid. In Fig.5 we show the best fit of these observed states to
our model.

\begin{center}
\includegraphics[scale=0.6,angle=0]{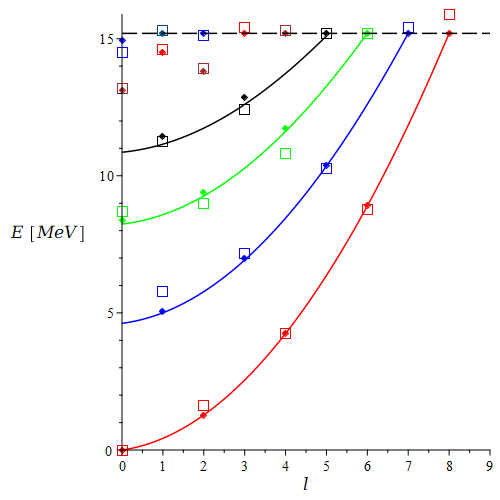}
\begin{center}
{{\bf Figure 5.} {Experimental energies (boxes), numerical bound
states (dots), and the anharmonic oscillator approximation (curves).}}
\end{center}
\end{center}

The dimensionless rotational constant $b_n$, given in (\ref{rotconst}),
converts to the physical value $B_n = 0.222 \, b_n$ MeV for the $n$-phonon
band. The rotational constants of the rotational bands in our model
are therefore $B_0 = 0.213 (0.212)$, $B_1 = 0.190 (0.137)$,
$B_2 = 0.170 (0.105)$, $B_3 = 0.146 (0.101)$ and $B_4 = 0.120$,
all in MeV. The bracketed quantities are the rotational
constants calculated in ref.\cite{BI} for the corresponding bands.
Although there are quantitative differences, there is a similar decrease
of $B_n$ as $n$ increases.

The detailed comparision between the experimental and model energies
in MeV, after the calibration, is as follows:
\begin{eqnarray}
& \mbox{Experiment} & \mbox{Model} \nonumber \\
\mbox{$0$-phonon band} & [0, 1.63, 4.25, 8.78, 15.87] 
& [0, 1.28, 4.25, 8.90, 15.19]  \nonumber \\
\mbox{$1$-phonon band} & [5.79, 7.16, 10.26, 15.37]
& [5.04, 6.97, 10.39, 15.19] \nonumber \\
\mbox{$2$-phonon band} & [8.7, 9.0, 10.8, 15.2]
& [8.36, 9.38, 11.71, 15.19] \nonumber \\
\mbox{$3$-phonon band} & [11.25, 12.4, 15.2] 
& [11.45, 12.86, 15.19] \nonumber \\
\mbox{$4$-phonon band} & [13.2, 13.9, 15.3]
& [13.11, 13.80, 15.19] \nonumber \\
\mbox{$5$-phonon band} & [14.6, 15.4] & [14.49, 15.19] \nonumber \\
\mbox{$6$-phonon band} & [14.5, 15.1] & [14.95, 15.19] \nonumber \\
\mbox{$7$-phonon band} & [15.3] & [15.19]. 
\end{eqnarray}

The TUNL table has no state very close to the model prediction of
14.95 MeV for the 6-phonon $0^+$ state. However,
a recent review of data from proton/Fluorine-19 scattering
experiments \cite{Lom} identifies some more $0^+$ states, including
one  at 14.9 MeV. This would better fit the 6-phonon band of our model.
There are also further $2^+$ states identified above 15 MeV, so there
is a 15.3 MeV alternative to our choice of the 15.1 MeV state as the
threshold 6-phonon state. If we had used these two states in our
final calibration, the threshold energy would have been slightly higher.

Identifying the 3-phonon ($n=3$) band of our model with
observed states is rather controversial. We find a good fit to the
energies and spin/parities of the observed $0^-$ band that Bijker and
Iachello describe using a combination of antisymmetric and symmetric
one-phonon oscillations of alpha-particles along the bipyramid
axis. In favour of our interpretation, we note that in earlier
work on rovibrational models for Oxygen-16 and Calcium-40 \cite{HKM,Cal},
it was found that 3-phonon states of a low-frequency
vibrational mode were needed to fit the experimental data. In
anharmonic models with a potential that flattens out, 3-phonon states
are not of very high energy, and cannot be ignored.

To confirm these band identifications, it would help to have predictions
for the frequencies of all vibrational modes of the bipyramid.
It would also help to clarify the status of the clear rotational
band described as a $1^-$ band in \cite{Tilley},
and interpreted this way in \cite{BI}. The problem is that the only
experimentally confirmed states in this band have spin/parities
$1^-, 3^-, 5^-, 7^-$, so it looks like a $0^-$ band.
We also note that some of the states assigned to the two
double-vibration $0^-$ bands in \cite{BI} are indicated in
\cite{Tilley} and in the TUNL table \cite{TUNL} to have isospin 1,
rather than the expected isospin 0. Greater clarity concerning the
odd spin, negative parity states of Neon-20 would be desirable.

It is of interest to relate the wormhole radius to the bipyramid
and two-cluster geometry more concretely using a simplified model.
Suppose that pointlike alpha-particles of mass $m_\alpha$ are located at
\bea
& \left(1,0,\quart z \right) \,,
\left(-\half, \frac{\sqrt{3}}{2}, \quart z \right) \,,
\left(-\half, -\frac{\sqrt{3}}{2}, \quart z \right) \,,
\left(0,0, \sqrt{2} + \quart z \right) \,, \nonumber \\
& \left(0,0, -\sqrt{2} - z \right) \,.
\eea
These form a tetrahedral four-alpha cluster centred at
$\left(0,0, \quart(\sqrt{2} + z) \right)$, accompanied by a single
alpha-particle at
$\left(0,0, -(\sqrt{2} + z) \right)$. The cluster separation is
$s = \frac{5}{4} (\sqrt{2} + z)$. When $z=0$ these five alphas form a
bipyramid of double-tetrahedron shape.

Let us now introduce a spatial scale factor $c$ for this structure.
The moment of inertia about any axis through the origin that is
orthogonal to the $x_3$-axis is found to be
\be
{\cal I} = \left( \frac{4}{5} s^2 + 3 \right) m_\alpha c^2 \,.
\ee
This can be identified with the moment of inertia defined by
the angular kinetic energy in (\ref{Lag}),
${\cal I} = \frac{4}{5} m_\alpha (r^2 + a^2) =
\frac{4}{5} (x^2 + 1) m_\alpha a^2$, where we have approximated $\mu$ as
$\frac{4}{5} m_\alpha$. Therefore
\be
\left( \frac{4}{5} s^2 + 3 \right) c^2 = \frac{4}{5} (x^2 + 1) a^2 \,.
\ee
When $z=0$ and $s = \frac{5}{4} \sqrt{2}$, then $x=0$, so the scale
factor is $c = \sqrt{\frac{8}{55}} \, a=2.07$ fm. The physical
cluster separation is $R = sc$, and a little further algebra gives
\be
R^2 = \left( x^2 + \frac{5}{11} \right) a^2 \,.
\ee
This relates the wormhole coordinate $x$ and wormhole radius $a$
to the physical geometry of the clusters. Note that $R$ is not
linearly related to $x$.

The separation of the clusters when they merge into the bipyramid is
$R = \sqrt{\frac{5}{11}} \, a = 3.65$ fm. This is the same as the separation 
of the alpha and Oxygen-16 clusters calculated by Zhou et al. \cite{Zho}.
The ground state wavefunctions extend from $x=0$ to approximately
$x =0.7$. This corresponds to a range of $R$-values between $3.65$ fm
and approximately $5.3$  fm. The 1-phonon wavefunctions have a peak at
about $x = 0.32$, corresponding to $R = 4.04$ fm, so the cluster
separation hardly exceeds that of the ground state wavefunctions. In
the 2-phonon wavefunctions, the node is at $x = 0.23$. The cluster
separation here is $R = 3.86$ fm, so the bipyramid is only slightly split.
On the other hand, the peak of the 2-phonon wavefunctions is around
$x = 0.54$. Here $R = 4.68$ fm, which is a substantially larger cluster
separation, larger than the sum of the root mean square radii of
an alpha-particle ($1.63$ fm) and an Oxygen-16 nucleus (2.72 fm). This large
cluster separation in the 2-phonon, higher nodal states is
in agreement with what is found using a variety of microscopic
cluster models, as illustrated in Fig.4.5 of ref.\cite{Fuj}. However,
recall that our wavefunctions are only defined outside the minimal
cluster separation, $3.65$ fm. 

\vspace{4mm}

\section{Conclusions}
\vspace{3mm}

We have proposed a model for the quantum states of Neon-20 interpolating
between the rotational excitations of the ground-state bipyramid, and
bound states of a separated alpha-particle and a four-alpha
cluster, i.e. an Oxygen-16 nucleus. The model combines a radial coordinate
$r$, related to the cluster separation via a moment of inertia,
and angular coordinates $(\theta, \phi)$ for the spatial orientation
of the axis joining the two clusters. $r=0$ at the bipyramid.

Our significant novel idea is that the geometry of the configuration
space is a 3-d spatial wormhole -- an Ellis--Bronnikov
wormhole -- with $SO(3)$ rotational symmetry. The $SO(3)$ orbit at
the wormhole throat at $r=0$ is not a point, as in Euclidean space,
but a 2-sphere of finite radius $a$, parametrising the orientation of
the bipyramid. On the wormhole configuration
space, we have added an attractive rotationally-symmetric,
short-range potential. This potential is proportional to the wormhole
curvature, and has a rather simple mathematical form.

The model's quantum Hamiltonian combines the (curved-space) Laplacian on the
wormhole together with the potential, and its bound states can be
classified into the rotational bands of $n$-phonon excitations of the
lowest-frequency vibrational mode of the bipyramid, which tends to
produce a $1+4$ cluster split.  

Our model differs from those based on Euclidean geometry, in that the
cluster separation is never less than what it is at the bipyramid. Also, 
the centrifugal repulsive potential for states with non-zero angular
momentum $l$ is not singular at the bipyramid, even though $r=0$.
Another property, arising from the choice of geometry and potential,
is that each rotational band has an angular momentum cut-off, and
therefore only a finite number of bound states. The threshold states
at the top of the bands all have the same energy, and form a sequence
of increasing angular momentum $l$, up to $l=8$ for our choice of
dimensionless parameter, $m=9$. The observed Neon-20 spectrum gives support for
this picture. The radial wavefunctions of the threshold states have
simple analytic forms, obtained by solving an associated
Legendre equation. This allows us to establish, by the Sturm oscillation
theorem, exactly how many lower-lying, true bound states there are, but
we have needed to find their energies numerically.

We have found that the model's energy spectrum matches the energies
of observed states of Neon-20 quite well, if we calibrate the energy
threshold to be at 15.19 MeV. The rotational bands include the
well-established $0_1^+$ ground state band, the lowest $0^-$ band, and the
$0^+_4$ band of broad, ``higher nodal'' states, and it supports the
assignment of observed $8^+$, $7^-$ and $6^+$ states between 15 and
16 MeV to these bands as threshold states. The calibration implies that
the wormhole's throat has radius $a = 5.42$ fm, corresponding to the
separation of the alpha-particle and Oxygen-16 cluster in the
ground-state bipyramid being 3.65 fm.

Unlike the harmonic oscillator potential, our potential (\ref{veff}),
shown in Fig.2, is a finite well, and it would be interesting to
analyse scattering states in this potential, to determine the widths of
resonant states along the lines of ref.\cite{BDV}, and to consider
alpha/Oxygen-16 fusion \cite{WN}. 

A challenge is to extend the model to include further vibrational
excitations of the bipyramid, and to understand theoretically all the
vibrational frequencies. The geometry of such an extended model will
be higher-dimensional and more complicated than a 3-d wormhole. 
It would also be interesting to investigate if spatial wormholes can model
the excitations of other nuclei that split asymmetrically into a pair
of clusters whose internal excitations can be neglected.

\vspace{3mm}

\section*{Acknowledgements}

NSM thanks David Jenkins for helpful discussions.

\end{document}